\documentclass[12pt]{iopart}

\usepackage{graphicx}
\usepackage{color}
\usepackage{siunitx}
\usepackage{graphicx}
\usepackage[normal]{subfigure}
\usepackage[usenames,dvipsnames]{xcolor}
\usepackage[ansinew]{inputenc}

\usepackage{framed}
\usepackage{caption}%[font=small,justification=justified,format=plain]{caption}
\usepackage{braket}

\begin{document}

\title{Quantum-dot based photonic quantum networks}

\author{Peter Lodahl}

\address{Niels Bohr Institute, University of Copenhagen, Blegdamsvej 17, DK-2100 Copenhagen, Denmark}
\ead{lodahl@nbi.ku.dk}
\vspace{10pt}
%\begin{indented}
%\item[]February 2014
%\end{indented}

\begin{abstract}
Quantum dots embedded in photonic nanostructures have in recent years proven to be a very powerful solid-state platform for quantum optics experiments. The combination of near-unity radiative coupling of a single quantum dot to a photonic mode and the ability to eliminate decoherence processes imply that an unprecedent light-matter interface can be obtained. As a result, high-cooperativity photon-emitter quantum interfaces can be constructed opening a path-way to deterministic photonic quantum gates for quantum-information processing applications. In the present manuscript, I review current state-of-the-art on quantum dot devices and their applications for quantum technology. The overarching long-term goal of the research field is to construct photonic quantum networks where remote entanglement can be distributed over long distances by photons.
\end{abstract}

%\maketitle

\section{Introduction}

The ultimate vision of photonic quantum technology is to construct a complex quantum network of stationary quantum nodes connected by flying photons in a fully quantum way, i.e. quantum entanglement may become distributed. Such a new photonic paradigm would have novel applications within secure quantum communication and is proposed as a way of scaling up quantum computers. It is popularly referred to as the 'quantum internet' \cite{Kimble2008QuantumInternet} and has been rooted in the atomic physics community  where an impressive proof-of-concept elementary two-node quantum network has been achieved \cite{Reiserer2015}. Solid-state alternatives to atomic single-photon emitters are attractive, since unlike atoms they do not require complex laser cooling and trapping techniques.  On the other hand solid-state systems are often considered to be 'noisy' in the sense that many potential decoherence processes may deteriorate quantum properties. Remarkably, self-assembled quantum dots (QDs) emitting single photons in the optical domain, notably InGaAs QDs embedded in GaAs semiconductors, have matured dramatically within the last few years. By systematically studying and combating the relevant decoherence processes \cite{Kuhlmann2013} impressive coherence has been demonstrated including near-perfect single-photon indistinguishability (above 98\%) of two subsequently emitted photons \cite{Somaschi16,Ding16} and transform-limited emission lines \cite{Kuhlmann2015}. Combined with the ability to dramatically enhance light-matter interaction in photonic nanostructures \cite{Lodahl2015RMP}, this enables near-deterministic single-photon sources with an internal collection efficiency exceeding 98 \% \cite{Arcari2014PRL}.
 %now becoming commercially available~\footnote{The company Sparrow Quantum has recently started commercializing highly-efficient nanophotonic waveguides containing quantum dot single-photon sources}.
 This major progress entails that QD single-photon sources are now gradually outperforming the traditional approaches based on atoms or spontaneous parametric down-conversion. QD sources benefit from fast operation speeds, excellent stability and brightness, as well as the potential scalability to multiple single photons and emitters. Notably, a number of impressive experiments have recently been implemented with QDs that have not been accomplished on other platforms. They include the experimental demonstration of a deterministic entangled cluster state of strings of photons \cite{Schwartz16}, boson sampling with so far five photons \cite{Wang16arxiv}, the detection of squeezed light correlations in resonance fluorescence \cite{Schulte15}, and the demonstration of an on-demand entangled photon source with higher than $90 \%$ fidelity \cite{Huber17}. Time is now to build on these and other achievements in order to scale QD photonic quantum technology and construct large and complex quantum architectures for quantum-information processing applications.

\begin{figure}[t]
\includegraphics[width=\columnwidth]{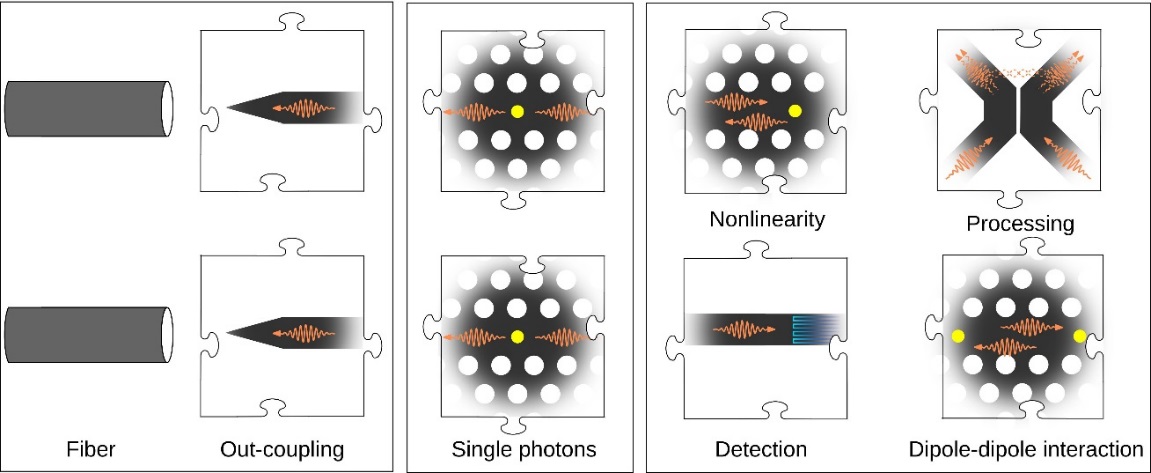}
\caption{\label{puzzle-pieces} Illustration of basic functionalities that can be implemented with QDs and photonic nanostructures. Left panel: efficient outcoupling taper sections from photonic nanostructures to optical fibers can be engineered to obtain highly efficient outcoupling of single photons. Center panel: a single QD in nanophotonic waveguide or cavity can be used as a highly efficient and coherent source of single photons. Two sources are shown in order to illustrate potential scalability of the approach. Right panel: a single QD efficiently coupled to a waveguide may be employed as a single-photon nonlinearity (upper left), complex photonic circuits may be constructed on a photonic chip for quantum processing of photons (upper right), single-photon superconducting detectors may be implemented on-chip for highly efficient detection (lower right), and multiple QDs may be coupled by engineering the dipole-dipole interaction in a nanophotonic waveguide. The figure is a courtesy of Sahand Mahmoodian.}
\end{figure}

The present manuscript reviews the current state-of-the-art on quantum photonics based on QD photon emitters towards the overarching goal of constructing photonic quantum networks. At present, a number of basic functionalities have been successfully demonstrated with generally impressive performance. These may constitute fundamental building blocks of photonic quantum networks. In that sense a current major challenge for theoretical quantum physicist is to develop resource efficient architectures tailored to the specific quantum hardware available, i.e., addressing how the basic 'puzzle pieces' can be combined. Figure ~\ref{puzzle-pieces} illustrates the 'puzzle pieces' already being developed for the QD platform, and that will be considered here. They include single-photon sources, single-photon nonlinearity, photonic circuitry, efficient-coupling to optical fibers, on-chip single-photon detectors, and multi-emitter coupling.

\begin{figure}[t]
\includegraphics[width=\columnwidth]{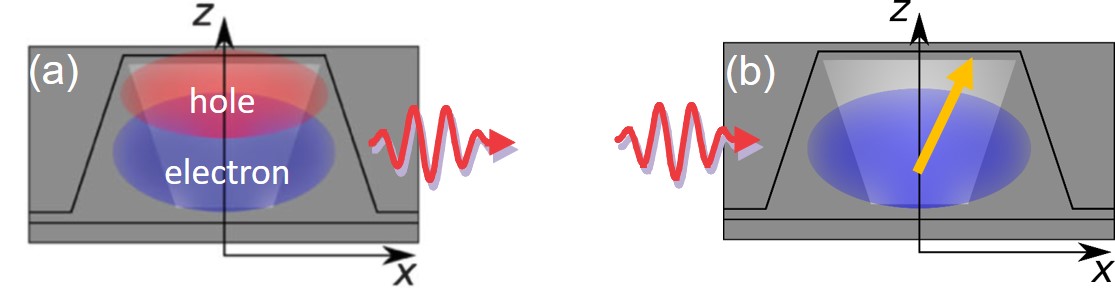}
\caption{\label{exciton+spin} Photon and spin qubits generated with QDs. The light grey areas illustrate the regions of InAs embedded in dark grey GaAs regions thereby creating an energy well potential for both electrons and holes. (a) An electron-hole pair trapped in the QD may recombine by emitting a single photon. (b) A single electron (or hole) trapped in the QD constitutes a spin qubit (orange arrow) that may be manipulated by optical methods.}
\end{figure}

QDs enable two different types of quantum resources: they may be employed as a source of single photons or alternatively a single spin trapped in a QD may be manipulated as a qubit. Figure ~\ref{exciton+spin} illustrates these two approaches. In the former case a single electron-hole pair is created in a QD by either optical excitation or by controllably tunneling carriers into the QD with the application of an electric field. The electron-hole pair subsequently recombines whereby a single photon is emitted by spontaneous emission within a lifetime of typically 1 ns for a QD in a bulk medium. In the latter approach a single carrier (electron or hole) is prepared in the QD by tunneling. The carrier spin is coupled to light since a photon may subsequently be absorbed by the QD creating a negatively (positively) charged exciton for the case of an initial electron (hole). Depending on the operational condition, the spin may have a coherence time of up to 1 $\mu$s (for holes) \cite{Greve10,Prechtel16}, which means that in principle many quantum operations can be carried out with fast optical control techniques before decoherence sets in. For some quantum-information applications, however, longer spin coherence times are required, e.g., in matter based quantum-repeater architectures. To this end, the QD platform must be interfaced with long-lived quantum memories based on, e.g., defect centers in diamond or atomic ensembles \cite{Hammerer10}. Such hybrid quantum network architectures are outside the scope of the present Review.

\section{Single-photon sources}

A QD efficiently coupled to a nanophotonic cavity or waveguide offers a promising route towards a deterministic single-photon source. Early pioneering work showed that Purcell enhancement in an optical cavity may be employed to both improve brightness and coherence of the source \cite{Santori02}. Ideally a useful single-photon source emits an optical pulse containing a single photon in a useful optical mode (e.g., into an optical fiber)  every time the QD is triggered by either an optical or electrical pulse.  Furthermore, most applications in quantum-information processing require that the emitted photons are fully coherent, which entails that decoherence processes occurring on a time scale of the radiative decay of the QD must be eliminated. Thus, three major figures-of-merit need to be optimized: i) the single-photon purity, ii) the single-photon coupling efficiency, and iii) the indistinguishability of the emitted photons.

\begin{figure}[t]
\includegraphics[width=\columnwidth]{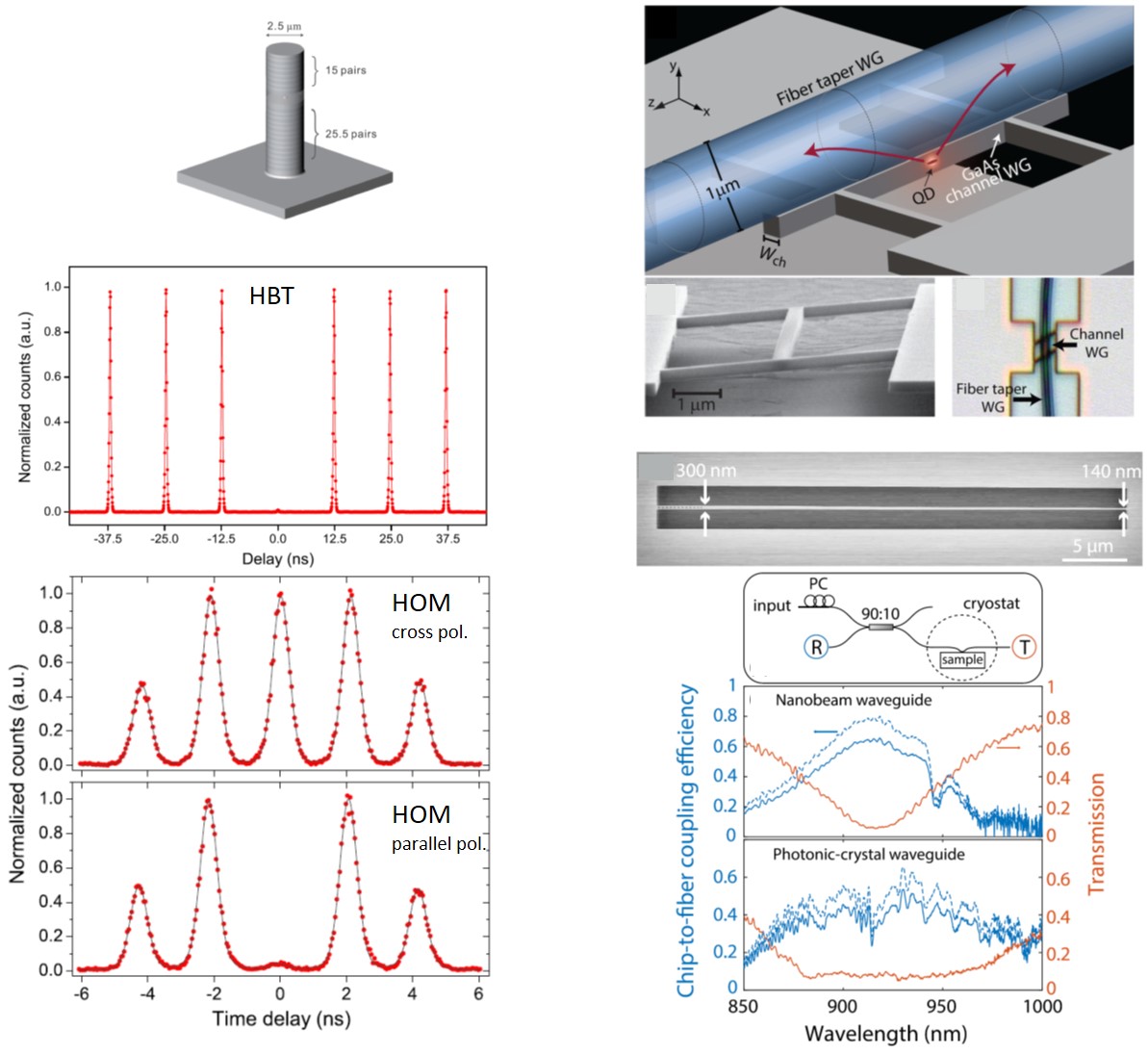}
\caption{\label{HBT-HOM-Efficiency} Purity, indistinguishability, and out-coupling efficiency of QD single-photon sources. Left column: Examples of HBT and Hong-Ou-Mandel (HOM) indistinguishability measurements by implementing resonant excitation on a QD in a micropillar cavity (upper figure shows the cavity structure). $g^{(2)}(0) = 0.009$ and a single-photon indistinguishability of $96.4 \%$ is extracted from the two sets of data. Figures reproduced from Ref. \cite{Ding16}. Right column: Examples of devices made for out-coupling single photon from a QD in a planar nanophotonic waveguide to an optical fiber with high efficiency by tailored evanescent coupling. The upper panel is reproduced from Ref. \cite{Davanco11}. Lower panel: application of this method to couple out single photons from high $\beta$-factor nanophotonic waveguides, i.e., nanobeam waveguides and photonic-crystal waveguides. By recording the reflection and transmission spectrum of the device, a chip-to-fiber coupling efficiency of $>80\%$ is obtained. Data reproduced from Ref. \cite{Daveau17}. }
\end{figure}

Re i): The discrete level structure of self-assembled QDs having widely separated optical transitions implies that they are capable of emitting high-purity single photons \cite{Michler00}. This is gauged in a Hanbury Brown - Twiss (HBT) correlation experiment by recording the multi-photon emission probability. By using samples with a low QD density and implementing (quasi)-resonant excitation to selectively excite only a single QD, excellent single-photon purity can be obtained at the level of $g^{(2)}(0) \leq  0.1 \%$, cf. Fig. \ref{HBT-HOM-Efficiency}.

Re ii): The overall coupling efficiency determines the brightness of the single-photon source and is comprised of several factors \cite{Lodahl2015RMP}: the excitation and single-photon emission probability of the QD, the efficiency with which the emitted photons are channeled to a single mode (the $\beta$-factor), and the transfer efficiency to a low-loss propagating mode, e.g., an optical fiber. Significant progress has been reported on all three engineering tasks: resonant $\pi$-pulse excitation allows deterministically preparing a single exciton in a QD \cite{He13} and electrically gated structures  eliminate blinking between different exciton complexes \cite{Warburton13}. Embedding QDs in photonic nanostructures allows reaching a near-unity $\beta$-factor, which has been obtained in nanophotonic waveguides \cite{Arcari2014PRL,Bleuse11} and cavities \cite{Somaschi16}. Finally, a variety of approaches and designs can be implemented for transferring the collected photons to an optical fiber using, e.g., tailored gratings for coupling light guided on planar structures vertically off the chip, or waveguide taper sections designed to couple directly to a fiber either by direct or evanescent coupling \cite{Davanco11,Tiecke15,Daveau17}.

Re iii): The indistinguishability of an emitted photon is determined by the amount of decoherence taking place within the emitter decay time. With a typical decay time of nanoseconds, the primary decoherence source is coupling of the QD to phonons and potentially photon jitter induced by relaxation processes originating from non-resonant excitation schemes \cite{Kiraz04}. The latter may be overcome by strict resonant excitation. Phonon interactions at the contrary are unavoidable and lead to two effects: wide phonon sidebands and residual broadening of the zero-phonon line. The former contribution (containing typically about 10 \% of the the total emission at low temperatures \cite{Lodahl2015RMP}) can readily be filtered away either by an external filter thereby reducing the overall efficiency or more favorably by implementing narrow-band Purcell enhancement. Hence the broadening of the zero-phonon line remains the fundamental decoherence mechanism. An in-depth theoretical study of this process including the role of the dimensionality of the photonic nanostructure can be found in Ref. \cite{Tighineanu17}. Importantly a clear route to indistinguishable photons with near-unity visibility can be laid out, which has also been confirmed experimentally \cite{Somaschi16,Ding16,Kirsanske17}.

\begin{figure}[t]
\includegraphics[width=\columnwidth]{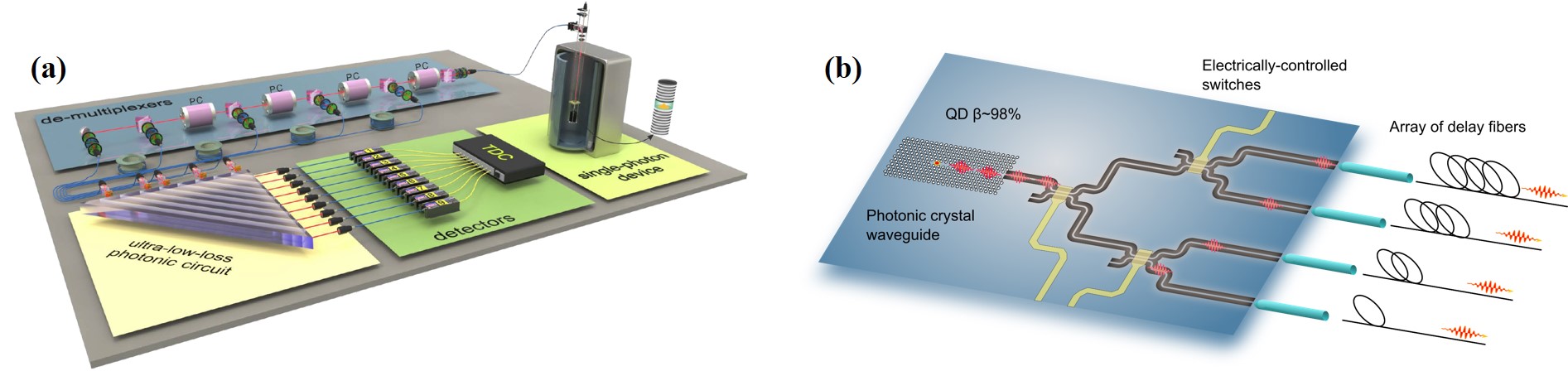}
\caption{\label{De-mux} Illustrations of schemes for de-multiplexing a single-photon source. (a) a five-photon  source obtained by cascading four Pockels cells (PCs) and polarizing beamsplitters to couple subsequently emitted photons from the QD into different optical fibers with different delay lengths. This de-multiplexed source is applied for proof-of-concept boson sampling. Figure reproduced from Ref. \cite{Wang16arxiv}. (b) Architecture of an integrated de-multiplexed source based on a QD coupled with high efficiency to a photonic-crystal waveguide. The emitted photons are transferred to dielectric waveguides and subsequently routed by electrical switches and coupled off-chip to different fiber delays in order to compensate for the time delay in-between photons. The figure is a courtesy of Leonardo Midolo. }
\end{figure}

The previous discussion illustrates that QDs are capable of producing a truly on-demand source of single photons by implementing, in a single device, the functionalities demonstrated so far only in different experiments. It is likely that such a source will be developed experimentally soon. In particular, it is remarkable that decoherence processes can be overcome in a solid-state system to such an extent that near-unity indistinguishability between subsequently emitted photons from the same QD can be obtained. Next step is to convert a single QD source generating photons as "pearls on a string" into a de-multiplexed source delivering many identical single photons simultaneously in individual modes. Such a source can be constructed by cascading electro-optical switches and coupling the photons into different optical fibers in order to overcome the time delay between subsequently emitted photons, cf. Fig. \ref{De-mux}(a). To this end, the demonstration of near-transform-limited QD emission lines \cite{Kuhlmann2015} has the remarkable consequence that a single QD can emit thousands of indistinguishable photons thereby providing a huge quantum resource. Indeed recent demonstration of indistinguishability exceeding 90 \% between two photons separated by more than 1000 emission events has proven this explicitly \cite{Wang16PRL}.

\section{Integrated quantum photonics}

As detailed in the previous section, currently the most reliable approach to generate multiple single photons utilizes a single integrated QD source where the photons are coupled off-chip to an optical fiber, and highly-efficient bulk optical components implemented for switching and routing, cf. Fig.~\ref{De-mux}(a). Nonetheless, the potential benefits of integrating functionalities on-chip are many in terms of stability, ease of operation, low loss, and speed \cite{OBrien09}. Figure~\ref{De-mux}(b) shows an architecture for an integrated de-multiplexed source of single photons based on a single QD in a planar nanophotonic waveguide platform. To implement this scheme requires the development of high-efficiency and fast switches, and multi-port waveguide-fiber interfaces. The former requires the development of new devices and functionalities on-chip where so-far the most common approach for reconfigurable circuitry has been the application of thermo-optic phase shifters \cite{Carolan15}. De-multiplexing requires fast switches, and two promising approaches include electro-optical modulation implemented directly in the host material of the QD (i.e. GaAs) \cite{Wang14,Midolo17}  or electro-mechanical coupling \cite{Midolo17rev}.

The crucial parameter determining the size of the photonic resource that can be generated with a de-multiplexed single-photon source is the overall efficiency, including switching, propagation, and coupling efficiencies. Consequently the rate of N-photon generation is $R_{\text{N}} = R_{\text{pump}} \times \eta^N/N$, where $\eta$ is the overall transmission efficiency from the source to the fiber and $R_{\text{pump}}$ is the repetition rate of the excitation laser. Exemplarily considering $\eta = 50 \%$ and $R_{\text{pump}} = 80 \: \text{MHz}$ (corresponding to the repetition rate of a Ti:Sapph. laser) corresponds to a 10-photon generation rate of $R_{10} \sim 8 \: \text{kHz},$ which illustrates the promising prospects of this approach while posing clear benchmarks for the efficiency of the applied switching and coupling technology.

Integrating a de-multiplexed source on a chip is one immediate goal. Another important task is to develop complex integrated tunable photonic circuits that can process the generated single photons leading to applications  for quantum simulations \cite{Guzik12}. Current state-of-the-art of this technology is a six-mode reprogrammable circuit capable of implementing universal high-fidelity quantum gates \cite{Carolan15}. Considerable effort is directed to the scaling of these circuits in working towards  a fully integrated system where photon source and processing circuit would be implemented on a single photonic chip. However, implementing on-chip the relatively long optical delays (typically tens of nanoseconds) required to interfere subsequently emitted photons, poses a challenge for this approach in requiring slow-light optical buffers. Consequently at present it seems most realistic to consider approaches based on two separate chips, connected by optical fibers, where one chip constitutes the source and the other chip the quantum processor.

The operation wavelength of a quantum photonics processor is an essential parameter. The preferred operation wavelength is within the telecom C-band $(1.55 \: \mu \text{m})$, where advanced high-performance optical components and fiber technology developed for telecom industry are present. The natural emission wavelength of self-assembled high quality QDs is around $950 \: \text{nm}$ and while significant efforts are taken to develop QDs in the telecom bands as well \cite{Kettler16,Kim16} the excellent performance in terms of coherence and efficiency required for quantum technology has not yet been achieved. An alternative approach employs frequency down-conversion of single photons from the near-infrared to the telecom band. Noise free conversion of single photons from a QD with an efficiency exceeding $30 \%$ has been demonstrated \cite{Kambs16}, which could be further increased by improving the in- and out-coupling efficiency through the nonlinear conversion crystal. The two-chip approach alluded to above has the asset that the frequency conversion step could naturally be incorporated in between the source chip and the processing chip. Furthermore, implementing frequency conversion of QD photons also has another advantage, since it provides a way of overcoming spectral inhomogeneities of photons emitted by different QDs by transducing them to the same wavelength in the C-band with the implementation of a tunable pump laser. This could prove a powerful way of scaling up QD systems to couple multiple emitter qubits (e.g., spins) in addition to the multiple photon qubits already considered.

Finally, another essential requirement is the ability to detect optical photons with very high efficiency. Very significant progress has been obtained in recent years with the development of superconducting nanowire single-photon detectors \cite{Hadfield09}, where the single-photon detection efficiency today approaches $100 \%$ and the speed is compatible with QD sources. Importantly these detectors can naturally be integrated on the planar GaAs platform containing QDs \cite{Reithmaier13}, which means that photon source, circuit, and detection could potentially be integrated on a single chip, which may pave the way to ultimately low-loss photonic quantum nodes.

\section{Single-photon nonlinearity}

The previous sections concerned the generation of single photons and their subsequent quantum interference in photonic circuits. Many quantum-information applications of photons require generating photon-photon interactions. Generally photons interact weakly meaning that standard nonlinearities are typically too weak to be operational at the level of single photons. However, the efficient coupling of single QDs to photonic nanostructures imply that such a system may be exploited to mediate a giant nonlinear response operational at the level of single photons. Two different approaches can be taken to reach a giant nonlinearity with a QD or any other quantum emitter: the QD can be strongly coupled to a cavity and the nonlinearity of the Jaynes-Cummings ladder exploited \cite{Faraon08,Reinhard12}, or a QD very efficiently coupled to a single optical mode in a cavity \cite{Bakker15,Bennett16} or waveguide \cite{Javadi15NatComm} may be used as a saturable nonlinearity. The latter approach is beneficial for ease of operation, since it is  less sensitive to the exact tuning of the wavelength of the QD. Furthermore, the saturable QD nonlinearity generally can be operated at a weaker incident photon flux \cite{Javadi15NatComm}. The key operational principle of the QD nonlinearity is as follows: a narrow-band pulse containing a single photon is reflected from a QD in a waveguide due to destructive interference of forward scattering and the incoming field. The efficient coupling (high $\beta$-factor) and coherent interaction, which can be achieved with QDs, implies that the interaction between the photon and the QD is deterministic. Since a two-level emitter can only scatter one photon at a time, two photons have an increased probability to be transmitted past the QD leading to a photon sorting process. From an experimental point of view this type of nonlinearity is attractive since no active control over the quantum state of the QD is required, i.e., the QD is merely exploited as a passive scatterer. On the other hand, the photon sorting is not ideal even for a perfect coupling efficiency. This implies that while a single-photon component of a pulse may be deterministically reflected, two- and higher-photon components are only partially transmitted and in this process entanglement is generated between the two photons \cite{Witthaut12EPL}. One approach of overcoming this limitation and obtain deterministic photon sorting has been suggested by combining the QD nonlinearity with a nonlinear spatio-temporal mode selector \cite{Ralph15PRL}. Such an approach may be applied for constructing a resource-efficient Bell-state analyzer \cite{Witthaut12EPL,Ralph15PRL}, cf. Section \ref{q-networks}. It should be mentioned that having a three-level emitter with coherent control of the level populations in the waveguide opens a range of additional opportunities. Such level schemes can be implemented with charged QDs, cf. the description in the next section. One exciting proposal is that of a photonic transistor controlled by only a single photon \cite{Chang07NatPhys}.

%\begin{figure}[t]
%\includegraphics[width=\columnwidth]{Figures/QD-nonlinearity.jpg}
%\caption{\label{QD-nonlinearity} Illustration of single-photon nonlinearity mediated by a QD efficiently coupled to a waveguide. Left: a narrow-band single photon pulse impinging on the QD is reflected with a large probability. Right: two photons are preferentially transmitted past the QD in this process becoming mutually entangled from the interaction with the QD.  }
%\end{figure}

\section{Spin-photon interfaces}

Introducing a single spin in a QD leads to a range of additional opportunities. It grants access to two separate ground states (spin up or down) that can be exploited as a quantum memory to encode a qubit. Either single holes or electrons can be employed and may be deterministically prepared by controlling the tunnel barriers in electrically contacted semiconductor structures. Various level structures can be implemented by the use of Zeeman tuning with a magnetic field, and spin initialization, manipulation, and read-out can be achieved rapidly with short optical pulses (typically in the nanosecond range), cf. Ref. \cite{Warburton13} for a detailed review of the physics of single spins in QDs. The spin lifetime can be in the range of milliseconds or longer in a large magnetic field, which is determined by phonon-mediated spin relaxation processes. However, in quantum applications the coherence time of the spin is the essential parameter. A coherence time of up to $T_2^* > 460 \: \mathrm{ns}$ has been reported for hole spins that has the advantage compared to electrons that they do not suffer from the Fermi contact hyperfine interaction with nuclear spins \cite{Prechtel16}. Importantly the spin coherence time can be several orders of magnitude longer than the spin manipulation time, which means that multiple quantum gate operations could be implemented within the spin coherence time. Unravelling spin decoherence mechanisms in QDs is a research topic in itself and is outside the scope of the present review. For a thorough discussion, see Refs. \cite{Warburton13,Urbaszek13}. It is anticipated that further research efforts will enable increasing the spin coherence time further. Nonetheless, it seems unlikely that it will be possible to extend the coherence time by many orders of magnitude in present QD systems. In particular, long-range quantum repeater protocols would require millisecond range quantum memories. This entails a  hybrid approach where the single photons generated by QDs are interfaced with long-lived quantum memories, e.g., vacancy centers in diamond \cite{Heshami14}, ions \cite{Meyer15}, or potentially mechanical systems.

\begin{figure}[t]
\includegraphics[width=\columnwidth]{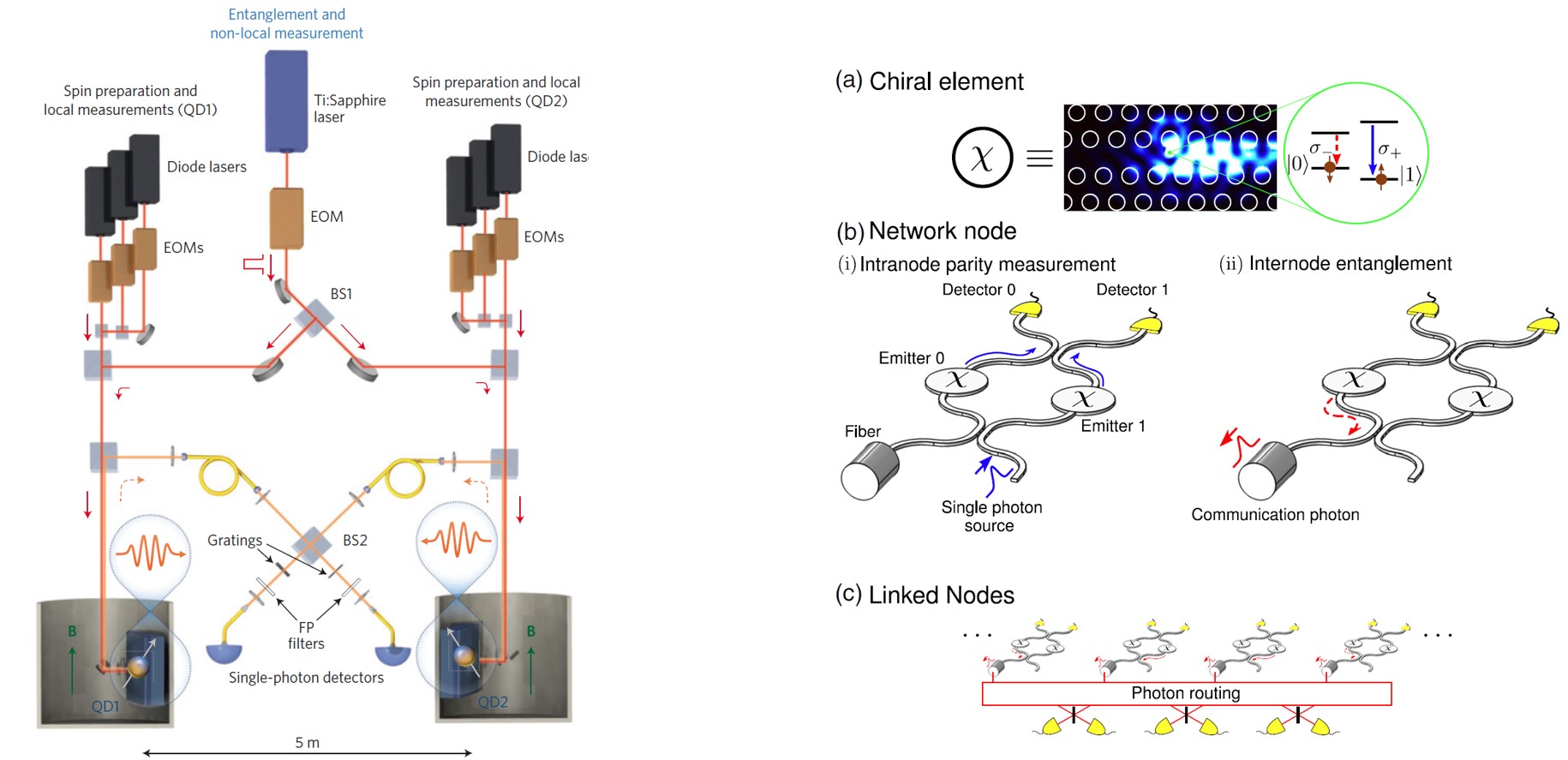}
\caption{\label{spin-applications} Left panel: experimental setup employed for the generation of remote entanglement between two QD hole spins in two different cryostats separated by $5 \: \mathrm{m}.$ Figure reproduced from Ref. \cite{Delteil16}. Right panel: Architecture of a photonic quantum network based on chiral photon-emitter elements (a) embedded in on-chip interferometers constituting fundamental quantum nodes (b). The quantum nodes are linked together by photons for remote entanglement distribution (c). Figure reproduced from Ref. \cite{Mahmoodian16}.     }
\end{figure}

 Considering a charged exciton in an external magnetic field, various four-level emitter schemes can be engineered with two ground states corresponding to either (pseudo)spin up or down of the electron (hole). The recombination of a charged exciton leads to the emission of polarized photons, where the polarization of the photon is correlated with the resulting spin state. In this way entanglement between the QD spin and the polarization of the emitted photon can be generated \cite{Greve10,Gao12} where in general hyper-entanglement between spin, polarization, and frequency is obtained. Subsequently the information in one of the variables can be erased to create a Bell state. This has been the resource enabling quantum teleportation between a single photon and a solid-state spin \cite{Gao13}. A further extension of this experiment was the demonstration of heralded entanglement between two spatially separate hole spins in QDs \cite{Delteil16}, cf. Fig. \ref{spin-applications} for the experimental layout.

 Another possibility is to couple QD spins directly to photon path by exploiting the concept of chiral light-matter interaction \cite{Lodahl17}. By tailoring the local electric field in a nanophotonic waveguide to match the in-plane circular dipole moment of one of the charged exciton transitions, it is possible to induce deterministic directional photon emission \cite{Sollner15}, i.e. $\sigma_{\pm}$ transitions emit in different directions. This effect was used to demonstrate an interface between QD spin and the photon path \cite{Coles16}. Based on this functionality a photonic quantum network architecture was put forward and its feasibility evaluated for QDs in nanophotonic waveguides \cite{Mahmoodian16}, cf. Fig. \ref{spin-applications}.

\section{Coupling of multiple quantum dots }

We have seen in previous sections how a single QD deterministically coupled to a single optical mode may be exploited as a source of multiple photonic qubits potentially generating a large quantum resource. It is natural to consider the option of scaling further up by coupling multiple QDs. This is generally a demanding task, as it is the case for any solid-state optical emitter, since inhomogeneous broadening of a QD ensemble (typically at the level of THz, which should be compared to the sub-GHz homogenous linewidth of a QD) implies that significant spectral mismatch would need to be overcome. Various tuning methods of QDs have been implemented including electric-field tuning \cite{Fry00} and strain tuning \cite{Ding10} that needs to be implemented locally to each individual QD. Eventually not only the central emission frequency of the different emitters must be controlled, but the complete photon pulses need to be matched in order to faithfully couple multiple QDs. An important figure-of-merit of the success is gauged in a quantum-interference experiment between two separate QD single-photon sources, where an interference visibility of $93 \%$ has been reported in a weak-excitation regime \cite{Stockill17}, which is a highly encouragin result for this approach.

Given the ability to radiatively couple two or more QDs, a photonic nanostructure can be constructed to enhance and tailor the coupling. Multiple QDs coupled by dipole-dipole interaction lead to the formation of collective quantum states that can have sub- or super-radiant character. In photonic nanostructures, the dipole-dipole interaction can be significantly extended beyond the behavior found in a homogenous (3D) medium, where the coupling falls off rapidly with distance $d$ (scales as $d^{-3}).$ In a 1D waveguide with radiative coupling to the waveguide as the dominating process, dipole-dipole interaction may be infinitely extended, i.e. in practice it is limited by the absorption or scattering loss of the guided mode. Controlling the distance between multiple emitters enables the switching between sub- and super-radiant behavior. Apart from the fundamental interest, the long-range dipole-dipole coupling may be exploited to construct quantum gates between stationary qubits \cite{Dzsotjan10}. Introducing chiral coupling opens new opportunities by the ability to engineer the directional coupling between emitters. This may lead to novel opportunities for quantum simulators and for entering new regimes of coupled photons and emitters. For instance directional coupling has been found to lead to the formation of quantum dimers in the steady state of the system \cite{Ramos14}. Another approach to scaling to multiple emitters exploits tunnel coupling between QDs grown in close mutual vicinity. With such an approach advanced level structures can be engineered in these QD molecules where it has been demonstrated that the spin coherence time can be significantly extended \cite{Weiss12}.

\section{Quantum network architectures}
\label{q-networks}
We have reviewed various basic quantum functionalities and hardware that can be implemented with the use of QDs embedded in modern nanophotonic structures. It is a pertinent task, to consider what applications and actual architectures can be implemented with these quantum resources and how to make them most resource efficient. This section briefly highlights some of the opportunities and future directions. A general observation, as discussed in detail above, is that QD systems can generate many single-photon qubits, while the scaling up to many matter qubits is expected to be much more challenging. It is important to have these general guidelines in mind when designing resource efficient quantum network architectures that is tailored to the quantum hardware based on QDs.

As discussed in the present Review, QDs in nanophotonic structures are very well suited for generating highly coherent single photons on demand at a high repetition rate. Such a source can generate thousands of indistinguishable photons \cite{Wang16PRL} that if efficiently routed could be an important photonic resource for quantum simulations \cite{Guzik12}. Furthermore, recent developments on how to efficiently fuse photonic qubits with linear optics have the renewed excitement on how to construct single-photon based quantum computing \cite{Rudolph17}.

\begin{figure}[t]
\includegraphics[width=\columnwidth]{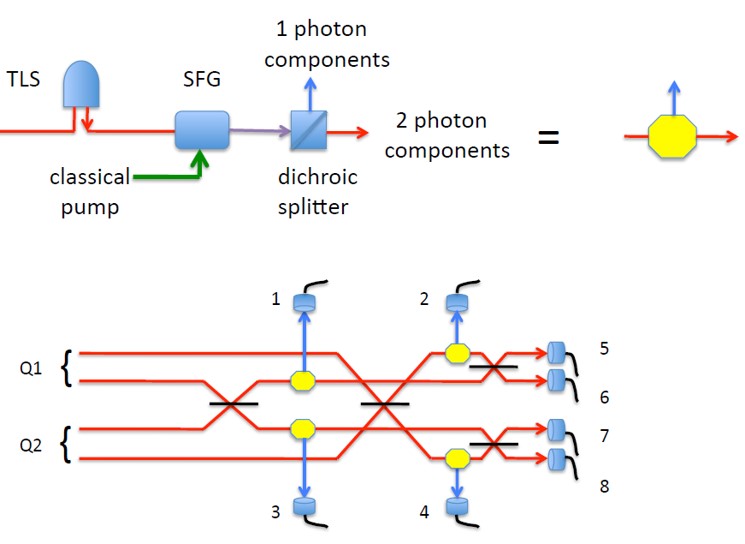}
\caption{\label{Bell-analyzer} Illustration of a deterministic Bell-state analyzer based on single-photon nonlinearity, frequency conversion, and linear optics. Upper plot: a deterministic two-level scattering (TLS) process induced, e.g., in a nanophotonic waveguide may be operated such that one and two photons are scattered to orthogonal spatio-temporal modes. The two modes can subsequently be selected by implementing active filtering, e.g. nonlinear sum-frequency generation (SFG) where the spatio-temporal mode of the classical pump is matched to, e.g., the one-photon component of the scattered light. Consequently, only the single-photon component is frequency converted and can be separated from the two-photon component by a dichroic splitter. Lower plot: optical circuit containing four photon-sorting elements that is capable of determining all four path-encoded Bell states by photon coincidence detection. Figure reproduced from Ref. \cite{Ralph15PRL}.}
\end{figure}

Introducing a single-photon nonlinearity to the toolbox opens for a range of additional possibilities. The most simple case is that of a two-level emitter, and here distortion of the optical pulses associated with photon-emitter bound state contributions has been interpreted as essentially a "no-go theorem" for quantum computation \cite{Shapiro06}. One work around, however, has been identified where the pulse distortion is actually exploited such that one- and two-photon pulses are scattered into orthogonal spectral-temporal modes that can subsequently be separated, e.g., by nonlinear frequency conversion \cite{Ralph15PRL}. With such a functionality all path-encoded photonic Bell states can be deterministically measured exploiting only four non-linear interaction process, linear optics, and photon detection, cf. Fig. \ref{Bell-analyzer}. Based on a Bell-state analyzer, quantum computing can in principle be implemented by teleportation based gates and linear optical quantum computing \cite{Gottesman99}. Alternatively, a photonic CZ gate can be directly based on the two-level emitter nonlinearity. However, in this case the pulse distortion introduced by photon scattering has to be undone by scattering an additional time on the emitter. However, since this scattering process is not time symmetric the photon pulse would need to be reverted in time in between the two scattering process. Such a time reversal of the photon pulses could in principle be obtained by storing, reverting, and releasing photon pulses in a gradient-echo quantum memory.

Having access to a 3-level $\Lambda$-scheme with two stable ground states leads to many more opportunities since now the emitter can store a qubit for extended times. Coherent control of the qubit state enables implementing a Duan-Kimble CNOT gate for photons where two subsequently interacting photons are entangled by their successive interaction with the same emitter \cite{Duan04}. Another related approach employs single-photon Raman interaction to deterministically swap the quantum state of the emitter and a photon \cite{Koshino10,Rosenblum17}. A full architecture of a quantum network based on QDs chirally coupled to nanophotonic waveguides has been put forward in Ref. \cite{Mahmoodian16}, cf. Fig. \ref{spin-applications}. Here the very large cooperativity obtainable with QDs leads to a gate fidelity approaching unity for experimentally realistic parameters.

Another highly exciting direction is to exploit a single QD efficiently coupled to a nanophotonic waveguide to directly produce entangled photons. To this end, Lindner and Rudolph have put forward the proposal of a photonic "cluster state machine gun" based on a QD that could potentially producing long strings of photons in an entangled cluster state \cite{Lindner09}. In this proposal, a single QD spin is brought into a superposition state of up and down and optically excited to a superposition of charged exciton states. Subsequently the QD decays by spontaneous emission of a single photon thereby creating an entangled state between the polarization of the emitted photon and the spin state. The excitation-emission process can be repeated multiple times in which case a large GHZ photonic state is generated. If $\pi/2$ rotations on the QD spin are carried out in-between photon emissions, an entangled photonic cluster state is created. A QD embedded in a nanophotonic waveguide could potentially serve as a source of large 1D cluster states, where the cluster size is determined by the number of emission events (determined by the QD lifetime that is typically $\sim 100 \: \mathrm{ps}$ in a nanophotonic structure) that can take place within the coherence time of the spin (up to $\sim \mu \mathrm{s}$). A pioneering experiment of photonic cluster state generation with a QD source has recently been reported \cite{Schwartz16}, which validates the approach.

Cluster states have attracted a lot of attention since it was realized that they enable universal quantum computing. The general philosophy behind such one-way quantum computing architectures \cite{Raussendorf03} is that a large scale entangled state is generated "up front" and computation  carried out by single-qubit operations induced by measurements. Importantly, quantum computing requires access to 2D photonic cluster states as opposed to the 1D cluster states considered above. Several proposals of how to achieve this with QDs have been put forward. One approach is to optically couple two QDs \cite{Economou10}, which requires local tuning of the two QDs into mutual resonance. An alternative proposal has been put forward based on a single quantum emitter, where the emitted photons are controllably coupled back to interact with the emitter and through this process generates the 2D cluster \cite{Pichler2017}. Such a "one-emitter quantum-computer architecture" is almost ideally suited for implementation based on QDs in nanophotonic structures. Another exciting potential applications of photonic cluster states would be in quantum repeater protocols without the necessity of a matter-based quantum memory \cite{Azuma15,Buterakos16}. Such an approach would potentially overcome the main bottleneck of the QD based platform for quantum networks resulting from the relatively short-lived quantum memory offered by the QD spin.

\section{Conclusions}

QDs in photonic nanostructures constitute an unprecedent photon-emitter interface at optical frequencies. The recent significant progress implies that advanced quantum-information processing protocols now become within experimental reach in the overall quest towards constructing photonic quantum networks. Such a distributed photonic quantum network would open new avenues for secure quantum communication and ultimately provide a way of scaling up quantum computers. It is currently an exciting and timely research topic to identify how the emerging quantum hardware can most efficiently be 'put to use' by laying out realistic and resource efficient quantum network architectures. Such a research programme requests a close interplay between theory and experiment in a continuous effort on how to exploit new quantum technology.

\section{Acknowledgements}
It is a pleasure to acknowledge all the excellent colleagues who have contributed to my understanding of solid-state photonic quantum networks  including A.S. S{\o}rensen, R.J. Warburton, T.C. Ralph, and A.G. White, as well as all past and present members of the Quantum Photonics Group at the Niels Bohr Institute, University of Copenhagen. I thank Nir Rotenberg for proof reading the manuscript. I gratefully acknowledge financial support from the European Research Council (ERC Advanced Grant \emph{SCALE}), Innovation Fund Denmark (Quantum Innovation Center \emph{Qubiz}), and the Danish Council for Independent Research.


\begin{thebibliography}{99}
\bibitem{Kimble2008QuantumInternet}
H.~J. Kimble,
 {\em The quantum internet, }
 Nature 453, 1023 (2008).

\bibitem{Reiserer2015}
A. Reiserer and G. Rempe,
{\em  Cavity-based quantum networks with single atoms and optical photons, }
Rev. Mod. Phys. 87, 1379 (2015).

\bibitem{Kuhlmann2013}
A.V. Kuhlmann, J. Houel, A. Ludwig, L. Greuter, D. Reuter, A.D. Wieck, M. Poggio, and R.J. Warburton,
{\em Charge noise and spin noise in a semiconductor quantum device, }
Nature Phys. 9, 570 (2013).

\bibitem{Somaschi16}
N. Somaschi, V. Giesz, L. De Santis, J. C. Loredo, M. P. Almeida, G. Hornecker, S. L. Portalupi, T. Grange, C. Anton, J. Demory, C. Gomez, I. Sagnes, N. D. Lanzillotti-Kimura, A. Lemaítre, A. Auffeves, A. G. White, L. Lanco, and P. Senellart,
{\em Near-optimal single-photon sources in the solid state, }
Nature Phot. 10, 340 (2016).


\bibitem{Ding16}
X. Ding, Y. He, Z.-C. Duan, N. Gregersen, M.-C. Chen, S. Unsleber, S. Maier, C. Schneider, M. Kamp, S. Hofling, C.-Y. Lu, and J.-W. Pan,
{\em On-demand single photons with high extraction efficiency and near-unity indistinguishability from a resonantly driven quantum dot in a micropillar, }
 Phys. Rev. Lett. 116, 020401 (2016).

\bibitem{Kuhlmann2015}
A.V. Kuhlmann, J.H. Prechtel, J. Houel, A. Ludwig, D. Reuter, A.D. Wieck, and R.J. Warburton,
{\em Transform-limited single photons from a single quantum dot,}
Nature Comm. 6, 8204 (2015)


\bibitem{Lodahl2015RMP}
P. Lodahl, S. Mahmoodian, and S. Stobbe,
 {\em Interfacing single photons and single quantum dots with photonic
  nanostructures, }
 Rev. Mod. Phys. 87, 347 (2015).

\bibitem{Arcari2014PRL}
M.~Arcari, I.~S\"ollner, A.~Javadi, S.~Lindskov Hansen, S.~Mahmoodian, J.~Liu,
  H.~Thyrrestrup, E.~H. Lee, J.~D. Song, S.~Stobbe, and P.~Lodahl,
 {\em Near-unity coupling efficiency of a quantum emitter to a photonic
  crystal waveguide, }
  Phys. Rev. Lett. 113, 093603 (2014).

\bibitem{Schwartz16}
I. Schwartz, D. Cogan, E.R. Schmidgall, Y. Don, L. Gantz, O. Kenneth, N.H. Lindner, and D. Gershoni,
{\em Deterministic generation of a cluster state of entangled photons,}
Science 354, 434 (2016).

\bibitem{Wang16arxiv}
H. Wang, Y. He, Y.-H. Li, Z.-E. Su, B. Li, H.-L. Huang, X. Ding, M.-C. Chen, C. Liu, J. Qin, J.-P. Li, Y.-M. He, C. Schneider, M. Kamp, C.-Z. Peng, S. Hoefling, C.-Y. Lu, and J.-W. Pan,
 {\em High-efficiency multiphoton boson sampling, }
Nature Phot. 11, 361 (2017).


\bibitem{Schulte15}
C.H.H. Schulte, J. Hansom, A.E. Jones, C. Matthiesen, C. Le Gall, and M. Atature,
{\em Quadrature squeezed photons from a two-level system,}
Nature 525, 222 (2015).


\bibitem{Huber17}
D. Huber, M. Reindl, Y. Huo, H. Huang, J.S. Wildmann, O.G. Schmidt, A. Rastelli, and R. Trotta,
{\em Highly indistinguishable and strongly entangled photons from symmetric GaAs quantum dots,}
Nature Comm. 8, 15506 (2017)

\bibitem{Greve10}
K. De Greve, P.L. McMahon, D. Press, T.D. Ladd, D. Bisping, C. Schneider, M. Kamp, L. Worschech, S. Hofling, A. Forchel, and Y. Yamamoto,
{\em Ultrafast coherent control and suppressed nuclear feedback of a single quantum dot hole qubit,}
Nature Physics 7, 872 (2011).

\bibitem{Prechtel16}
J.H. Prechtel, A.V. Kuhlmann, J. Houel, A. Ludwig, S.R. Valentin, A.D. Wieck, and R.J. Warburton,
{\em Decoupling a hole spin qubit from the nuclear spins, }
Nature Materials 15, 981 (2016)

\bibitem{Hammerer10}
K. Hammerer, A.S. S{\o}rensen, and E.S. Polzik,
{\em Quantum interface between light and atomic ensembles,}
Rev. Mod. Phys. 82, 1041 (2010).

\bibitem{Santori02}
C. Santori, D. Fattal, J. Vukovic, G.S. Solomon, and Y. Yamamoto,
{\em Indistinguishable photons from a single-photon device,}
Nature 419, 594 (2002).

\bibitem{Michler00}
P. Michler, A. Kiraz, C. Becher, W.V. Schoenfeld, P.M. Petroff, L. Zhang, E. Hu, and A. Imamoglu,
{\em A quantum dot single-photon turnstile device, }
Science 290, 2282 (2000).

\bibitem{He13}
Y.-M. He, Y. He, Y.-J. Wei, D. Wu, M. Atature, C. Schneider, S. Hofling, M. Kamp, C.-Y. Lu, and J.-W. Pan,
{\em On-demand semiconductor single-photon source with near-unity indistinguishability,}
Nature Nano. 8, 213 (2013).

\bibitem{Warburton13}
R.J. Warburton,
{\em Single spins in self-assembled quantum dots,}
Nature Materials 12, 483 (2013).

\bibitem{Bleuse11}
J. Bleuse, J. Claudon, M. Creasey, N.S. Malik, J.-M. Gerard, I. Maksymov, J.-P. Hugonin, and P. Lalanne
{\em Inhibition, enhancement, and control of spontaneous emission in photonic nanowires}
Phys. Rev. Lett. 106, 103601 (2011).


\bibitem{Davanco11}
M. Davanco, M.T. Rakher, W. Wegscheider, D. Schuh, A. Badolato, and K. Srinivasan,
{\em Efficient quantum dot single photon extraction into an optical fiber using a nanophotonic directional coupler, }
Appl. Phys. Lett. 99, 121101 (2011).


\bibitem{Tiecke15}
T.G. Tiecke, K.P. Nayak, J.D. Thompson, T. Peyronel, N.P. de Leon, V. Vuletic, and M.D. Lukin,
{\em Efficient fiber-optical interface for nanophotonic devices,}
Optica 2, 70 (2015).


\bibitem{Daveau17}
R.S. Daveau, K.C. Balram, T. Pregnolato, J. Liu, E.H. Lee, J.D. Song, V. Verma, R. Mirin, S. Woo Nam, L. Midolo, S. Stobbe, K. Srinivasan, and P. Lodahl,
{\em Efficient fiber-coupled single-photon source based on quantum dots in a photonic-crystal waveguide, }
Optica 4, 178 (2017).


\bibitem{Kiraz04}
A. Kiraz, M. Atature, and A. Imamoglu,
{\em Quantum-dot single-photon sources: Prospects for applications in linear optics
quantum-information processing,}
Phys. Rev. A 69, 032305 (2004).

\bibitem{Tighineanu17}
P. Tighineanu, C.L. Dreessen, C. Flindt, P. Lodahl, and A.S. S{\o}rensen,
{\em Phonon decoherence of quantum dots in photonic structures: Broadening of the zero-phonon line and the role of dimensionality,}
arXiv:1702.04812 (2017).

\bibitem{Kirsanske17}
G. Kirsanske, H. Thyrrestrup, R.S. Daveau, C.L. Dreessen, T. Pregnolato, L. Midolo, P. Tighineanu, A. Javadi, S. Stobbe, R. Schott, A. Ludwig, A.D. Wieck, S.I Park, J.D. Song, A.V. Kuhlmann, I. Sollner, M.C. Lobl, R.J. Warburton, and P. Lodahl,
{\em Indistinguishable and efficient single photons from a quantum dot in a planar nanobeam waveguide,}
arXiv:1701.08131 (2017).

\bibitem{Wang16PRL}
H. Wang, Z.-C. Duan, Y.-H. Li, S. Chen, J.-P. Li, Y.-M. He, M.-C. Chen, Y. He, X. Ding, C.-Z. Peng, C. Schneider, M. Kamp, S. Hofling, C.-Y. Lu, and J.-W. Pan,
{\em Near-transform-limited single photons from an efficient solid-state quantum emitter, }
Phys. Rev. Lett. 116, 213601 (2016).

\bibitem{OBrien09}
J.L. O'Brien, A. Furusawa, and J. Vuckovic,
{\em Photonic quantum technologies,}
Nature Photonics 3, 687 (2009).


\bibitem{Carolan15}
J. Carolan, C. Harrold, C. Sparrow, E. Martin-Lopez, N.J. Russell, J.W. Silverstone, P.J. Shadbolt, N. Matsuda, M. Oguma, M. Itoh, G.D. Marshall, M.G. Thompson, J.C.F. Matthews, T. Hashimoto, J.L. O'Brien, and A. Laing,
{\em Universal linear optics,}
Science 349, 711 (2015).

\bibitem{Wang14}
 J. Wang, A. Santamato, P. Jiang, D. Bonneau, E. Engin, J.W. Silverstone, M. Lermer, J. Beetz, M. Kamp, S. Hofling, M.G. Tanner, C.M. Natarajan, R.H. Hadfield, S.N. Dorenbos, V. Zwiller, J.L. O'Brien, and M.G. Thompson,
 {\em Gallium arsenide (GaAs) quantum photonic waveguide circuits,}
 Opt. Commun. 327, 49 (2014).

\bibitem{Midolo17}
L. Midolo et al., in preparation (2017).


\bibitem{Midolo17rev}
L. Midolo, A. Schliesser, and A. Fiore, in preparation (2017).

\bibitem{Guzik12}
A. Aspuru-Guzik and P. Walther,
{\em Photonic quantum simulators}
Nature Physics 8, 285 (2012).

\bibitem{Kettler16}
J. Kettler, M. Paul, F. Olbrich, K. Zeuner, M. Jetter, and P. Michler,
{\em Single-photon and photon pair emission from MOVPE-grown In(Ga)As quantum dots: shifting the emission wavelength from 1.0 to 1.3 $\mu$m }
Appl. Phys. B 122, 48 (2016).

\bibitem{Kim16}
J.-H. Kim, T. Cai, C.J.K. Richardson, R.P. Leavitt, and E. Waks,
{\em Two-photon interference from a bright single-photon source at telecom wavelengths, }
Optica 3, 577 (2016).

\bibitem{Kambs16}
B. Kambs, J. Kettler, M. Bock, J.N. Becker, C. Arend, A. Lenhard, S.L. Portalupi, M. Jetter, P. Michler, and C. Becher,
{\em Low-noise quantum frequency down-conversion of indistinguishable photons,}
Opt. Express 24, 22250 (2016).

\bibitem{Hadfield09}
R.H. Hadfield,
{\em Single-photon detectors for optical quantum information applications,}
Nature Photonics 3, 696 (2009).

\bibitem{Reithmaier13}
G. Reithmaier, S. Lichtmannecker, T. Reichert, P. Hasch, K. Muller, M. Bichler, R. Gross, and J.J. Finley,
{\em On-chip time resolved detection of quantum dot emission using integrated superconducting single photon detectors,}
Sci. Reports 3, 1901 (2013).

\bibitem{Faraon08}
A. Faraon, I. Fushman, D. Englund, N. Stoltz, P. Petroff, and J. Vukovic,
{\em Coherent generation of non-classical light on a chip via photon-induced tunnelling and blockade,}
Nature Phys. 4, 859 (2008).

\bibitem{Reinhard12}
A. Reinhard, T. Volz, M. Winger, A. Badolato, K.J. Hennessy, E.L. Hu, and A. Imamoglu,
{\em Strongly correlated photons on a chip,}
Nature Phot. 6, 93 (2012).

\bibitem{Bakker15}
M.P. Bakker, T. Ruytenberg, W. Loffler, A. Barve, L. Coldren, M.P. van Exter, and D. Bouwmeester,
{\em Quantum dot nonlinearity through cavity-enhanced feedback with a charge memory}
Phys. Rev. B 91, 241305(R) (2015)

\bibitem{Bennett16}
A.J. Bennett, J.P. Lee, D.J.P. Ellis, I. Farrer, D.A. Ritchie, and A.J. Shields,
{\em A semiconductor photon-sorter}
Nature Nano. 11, 857 (2016).

\bibitem{Javadi15NatComm}
A. Javadi, I. S\"{o}llner, M. Arcari, S.L. Hansen, L. Midolo, S. Mahmoodian, G. Kirsanske, T. Pregnolato, E.H. Lee, J.D. Song, S. Stobbe, and P. Lodahl,
{\em Single-photon non-linear optics with a quantum dot in a waveguide,} Nature Comm. 6, 8655 (2015)

\bibitem{Witthaut12EPL}
D. Witthaut, M.D. Lukin, and A.S. S{\o}rensen,
{\em Photon sorters and QND detectors using single photon emitters,}
Europhys. Lett. 97, 50007 (2012).

\bibitem{Ralph15PRL}
T.C. Ralph, I. S\"{o}llner, S. Mahmoodian, A.G. White, and P. Lodahl,
{\em Photon sorting, efficient Bell measurements and deterministic CZ gate using a passive two-level nonlinearity, }
Phys. Rev. Lett. 114, 173603 (2015).

\bibitem{Chang07NatPhys}
D.E. Chang, A.S. S{\o}rensen, E.A. Demler, and M.D. Lukin,
{\em A single-photon transistor using nanoscale surface plasmons,}
Nature Phys. 3, 807 (2007).

\bibitem{Urbaszek13}
B. Urbaszek, X. Marie, T. Amand, O. Krebs, P. Voisin, P. Maletinsky, A. Hogele, and A. Imamoglu,
{\em Nuclear spin physics in quantum dots: An optical investigation,}
Rev. Mod. Phys. 85, 79 (2013).

\bibitem{Heshami14}
K. Heshami, C. Santori, B. Khanaliloo, C. Healey, V.M. Acosta, P.E. Barclay, and C. Simon,
{\em Raman quantum memory based on an ensemble of nitrogen-vacancy centers coupled to a microcavity,}
Phys. Rev. A 89, 040301(R) (2014).

\bibitem{Meyer15}
H.M. Meyer, R. Stockill, M. Steiner, C. Le Gall, C. Matthiesen, E. Clarke, A. Ludwig, J. Reichel, M. Atature, and M. Kohl,
{\em Direct photonic coupling of a semiconductor quantum dot and a trapped ion,}
Phys. Rev. Lett. 114, 123001 (2015).

\bibitem{Gao12}
W.B. Gao, P. Fallahi, E. Togan, J. Miguel-Sanchez, and A. Imamoglu,
{\em Observation of entanglement between a quantum dot spin and a single photon,}
Nature 491, 426 (2012).

\bibitem{Gao13}
W.B. Gao, P. Fallahi, E. Togan, A. Delteil, Y.S. Chin, J. Miguel-Sanchez, and A. Imamoglu,
{\em Quantum teleportation from a propagating photon to a solid-state spin qubit,}
Nature Comm.  4, 2744 (2013).

\bibitem{Delteil16}
A. Delteil, Z. Sun, W.B. Gao, E. Togan, S. Faelt, and A. Imamoglu,
{\em Generation of heralded entanglement between distant hole spins,}
Nature Phys. 12, 218 (2016).

\bibitem{Lodahl17}
P. Lodahl, S. Mahmoodian, S. Stobbe, A. Rauschenbeutel, P. Schneeweiss, J. Volz, H. Pichler, and P. Zoller,
{\em Chiral quantum optics,}
Nature 541, 473 (2017).

\bibitem{Sollner15}
I. S\"{o}llner, S. Mahmoodian, S. Lindskov Hansen, L. Midolo, A. Javadi, G. Kirsanske, T. Pregnolato, H. El-Ella, E.-H. Lee, J.D. Song, S. Stobbe, and P. Lodahl,
{\em Deterministic photon-emitter coupling in chiral photonic circuits,}
Nature Nano. 10, 775 (2015).

\bibitem{Coles16}
R.J. Coles, D.M. Price, J.E. Dixon, B. Royall, E. Clarke, P. Kok, M.S. Skolnick, A.M. Fox, and M.N. Makhonin,
{\em Chirality of nanophotonic waveguide with embedded quantum emitter for unidirectional spin transfer,}
Nature Comm. 7, 11183 (2016).

\bibitem{Mahmoodian16}
S. Mahmoodian, P. Lodahl, and A. S. S{\o}rensen,
{\em Quantum networks with chiral light-matter interaction in waveguides,}
Phys. Rev. Lett. 117, 240501 (2016).

\bibitem{Fry00}
P.W. Fry, I.E. Itskevich, D.J. Mowbray, M.S. Skolnick, J.J. Finley, J.A. Barker, E.P. O'Reilly, L.R. Wilson, I.A. Larkin, P.A. Maksym, M. Hopkinson, M. Al-Khafaji, J.P.R. David, A.G. Cullis, G. Hill, and J.C. Clark,
{\em Inverted electron-hole alignment in InAs-GaAs self-assembled quantum dots,}
Phys. Rev. Lett. 84, 733 (2000).

\bibitem{Ding10}
F. Ding, R. Singh, J.D. Plumhof, T. Zander, V. Krapek, Y.H. Chen, M. Benyoucef, V. Zwiller, K. Dorr, G. Bester, A. Rastelli, and O.G. Schmidt,
{\em Tuning the exciton binding energies in single self-assembled InGaAs/GaAs quantum dots by piezoelectric-induced biaxial stress,}
Phys. Rev. Lett. 104, 067405 (2010).

\bibitem{Stockill17}
R. Stockill, M.J. Stanley, L. Huthmacher, E. Clarke, M. Hugues, A.J. Miller, C. Matthiesen, C. Le Gall, and M. Atature,
{\em Phase-tuned entangled state generation between distant spin qubits,}
arXiv:1702.03422 (2017).

\bibitem{Dzsotjan10}
D. Dzsotjan, A.S. S{\o}rensen, and M. Fleischhauer,
{\em Quantum emitters coupled to surface plasmons of a nanowire: A Greens function approach,}
Phys. Rev. B 82, 075427 (2010).

\bibitem{Ramos14}
T. Ramos, H. Pichler, A.J. Daley, and P. Zoller,
{\em Quantum spin dimers from chiral dissipation in cold-atom chains,}
Phys. Rev. Lett. 113, 237203 (2014).

\bibitem{Weiss12}
K.M. Weiss, J.M. Elzerman, Y.L. Delley, J. Miguel-Sanchez, and A. Imamoglu,
{\em Coherent two-electron spin qubits in an optically active pair of coupled InGaAs quantum dots,}
Phys. Rev. Lett. 109, 107401 (2012).

\bibitem{Rudolph17}
T. Rudolph, {\em Why I am optimistic about the silicon-photonic route to quantum computing,}
APL Photonics 2, 030901 (2017).

\bibitem{Shapiro06}
J.H. Shapiro,
{\em Single-photon Kerr nonlinearities do not help quantum computation,}
Phys. Rev. A 73, 062305 (2006).

\bibitem{Gottesman99}
D. Gottesman and I.L. Chuang,
{\em Demonstrating the viability of universal quantum computation using teleportation and single-qubit operations,}
Nature 402, 390 (1999).

\bibitem{Duan04}
L.M. Duan and H.J. Kimble,
{\em Scalable photonic quantum computation through cavity-assisted interactions,}
Phys. Rev. Lett. 92, 127902 (2004).

\bibitem{Koshino10}
K. Koshino, S. Ishizaka, and Y. Nakamura,
{Deterministic photon-photon $\sqrt{\mathrm{SWAP}}$ gate using a $\Lambda$ system,}
Phys. Rev. A 82, 010301(R) (2010).

\bibitem{Rosenblum17}
S. Rosenblum, A. Borne, and B. Dayan,
{\em Analysis of deterministic swapping of photonic and atomic states through single-photon Raman interaction,}
Phys. Rev. A 95, 033814 (2017).

\bibitem{Lindner09}
N.H. Lindner and T. Rudolph,
{\em Proposal for pulsed on-demand sources of photonic cluster state strings,}
Phys. Rev. Lett. 103, 113602 (2009).

\bibitem{Raussendorf03}
R. Raussendorf, D.E. Browne, and H.J. Briegel,
{\em Measurement-based quantum computation on cluster states,}
Phys. Rev. A 68, 022312 (2003).

\bibitem{Economou10}
S.E. Economou, N. Lindner, and T. Rudolph, {\em Optically generated 2-dimensional photonic cluster state from coupled quantum dots,}
Phys. Rev. Lett. 105, 093601 (2010).

\bibitem{Pichler2017}
H. Pichler, S. Choi, P. Zoller, and M.D. Lukin, {\em Photonic tensor networks produced by a single quantum emitter, }
arXiv:1702.02119

\bibitem{Azuma15}
K. T. K. Azuma and H.-K. Lo,
{\em All-photonic quantum repeaters,}
Nat. Commun. 6, 6787 (2015).

\bibitem{Buterakos16}
D. Buterakos, E. Barnes, and S.E. Economou,
{Deterministic generation of all-photonic quantum repeaters from solid-state emitters,}
arXiv:1612.03869 (2016).








\end{thebibliography}
\end{document}